\documentclass[epj,twocolumn]{webofc}
\usepackage[varg]{txfonts}

\woctitle{ISVHECRI}

\begin{document}
\title{pp interaction in extended air showers}
%
%

\author{A Kendi Kohara\inst{1} {\email{kendi@if.uftj.br}} 
 \and
  Erasmo Ferreira\inst{1} {\email{erasmo@if.ufrj.br  }}
 \and
  Takeshi Kodama \inst{1} {\email{tkodama@if.ufrj.br}}
 }
\institute{Instituto de F\'isica, Universidade Federal do Rio de Janeiro,
C.P. 68528, Rio de Janeiro 21945-970, RJ, Brazil      }

\abstract{ 

Applying the recently constructed analytic representation for the pp 
scattering amplitudes, we present a study of p-air cross sections, with comparison 
to the data from  Extensive Air Shower (EAS) measurements.  The amplitudes 
describe with  precision  all available  accelerator data  at ISR, SPS 
and LHC  energies, and its theoretical basis, together with the very smooth
energy dependence of parameters controlled by unitarity and dispersion relations, 
permit reliable extrapolation to higher energies and to asymptotic ranges. 
The comparison with cosmic ray data is very satisfactory in the whole pp energy 
interval from 1 to 100 TeV. High energy asymptotic behaviour  
of cross sections is  investigated in view of the geometric scaling property  of 
the amplitudes. The amplitudes predict that the proton does not behave as a black 
disk even at asymptotically  high enegies, and we discuss possible non-trivial 
consequences of this fact for pA collision cross sections at higher energies. }

\maketitle

\section{Motivation}

\label{intro}

The data on pA collisions extracted from studies of Extensive Air Showers
(EAS) reaches the 100 TeV region of pp center of mass energies ($\sqrt{s}$%
) that are well above the present accelerator experiments. These data
contain very important information on the dynamics of strong
interactions at extreme high energies and also on the origin of high energy
cosmic rays (CR) . Thus the earth atmosphere is used as a detector
of high energy particles to investigate the high energy phenomena in the
Universe. On the other hand, different from the laboratory experiment,
analysis of these data is more dependent on  models that permit reliable 
extrapolation to this  energy region.  

Recent  detailed analyses \cite{KEK_2013,KEK_2013b, KEK-JPhysG2014, KEK-to-be,
KEK_2014}  
of the experimental pp and $\mathrm{p\bar{p}}$
scattering data based on a QCD inspired model \cite{dosch,ferreira1} 
reproduces the  observed cross sections for all energies with high accuracy.
These analyses lead to a precise analytic representation of the elastic 
scattering amplitudes, 
disentangling the real and imaginary parts, respecting  unitarity and
causality conditions, with regular  energy dependence  that  
  gives confidence in extrapolation to higher energies.
The purpose of the present work is to calculate the proton-air production
cross section in the framework of Glauber model using this representation of
pp scattering as input, and compare the results to the experimental values
obtained from the available cosmic ray (CR) data. We are mainly concerned with
the energies beyond the LHC experiments but also present results for EAS
experiments in the region below 1 TeV.

Our analysis of energy dependence of amplitudes and observables in pp
collisions shows that the total cross section has a neat $\log ^{2}{s}$ form 
\cite{KEK_2014}, as already indicated in several analyses \cite{PDG}. An
important new feature   is that the slope parameters, 
 $B_{I}$ and $B_{R}$, also have a $\log ^{2}{s}$ dependence. This is new  
  finding, since the generally accepted idea is that the slope of the
differential cross sections varies like simple linear $\log {s}$, as in
Regge phenomenology. This result has a crucial effect for the use of
Glauber formalism in the analysis of p-air extended showers at the high
energies of our concern, since the value of the slope $B_{I}$, together with
the value of the total cross section, are the basic  
inputs of the calculation.

For the application of the standard Glauber approach, we basically need the
amplitudes in forward scattering. In these conditions our amplitudes take
simpler exponential forms requiring only two parameters to specify each
amplitude. The relevant parameters are then the total cross section 
$\sigma $, the ratio $\rho $ between real and imaginary parts at $t=0$, and the
slopes $B_{I}$ and $B_{R}$ of each of the two parts. Our full-$t$ analysis 
\cite{KEK_2014} provides the energy dependence of these quantities with
simple analytical forms that are appropriate for the whole energy range from
50 GeV to 100 TeV.

On the other hand, the $\log ^{2}{s}$ dependence of the slopes is intimately
related to the unitarity condition, as can be seen more clearly in the 
$b$-space representation. Another interesting consequences of our
representation is that a proton does not behave as a black disk even in
asymptotic energy domain. This fact also has important
consequences to the estimate of pA collision cross section in the ultra-high
energy domain.

\section{Forward Amplitudes and Observables \label{forward}}

In the treatment of elastic pp and p$\mathrm{{\bar{p}}}$ scattering in the
forward direction, with amplitudes approximated by pure exponential forms,
the differential cross section is written 
\begin{eqnarray}
\frac{d\sigma }{dt} &=&\pi \left( \hbar c\right) ^{2}~\Big\{\Big[\frac{\rho
\sigma }{4\pi \left( \hbar c\right) ^{2}}~{{e}^{B_{R}t/2}+F^{C}(t)\cos {%
(\alpha \Phi )}\Big]^{2}}  \nonumber \\
&&+\Big[\frac{\sigma }{4\pi \left( \hbar c\right) ^{2}}~{{e}%
^{B_{I}t/2}+F^{C}(t)\sin {(\alpha \Phi )}\Big]^{2}\Big\}~,}
\label{diffcross_eq}
\end{eqnarray}
where $t\equiv -|t|$ and we must allow different values for the slopes $%
B_{I} $ and $B_{R}$ of the imaginary and real amplitudes. With $\sigma$ in
milibarns and $|t|$ in GeV$^2$, we have $\left( \hbar c\right)^2~= ~0.3894$.
Since we work with $B_{R}\neq B_{I}$ , treatment of the Coulomb interference
requires a more general expression for the Coulomb phase, which has been
developed before \cite{KEK_2013}. However, in the present work we only need
the forward ($|t|=0$ ) nuclear amplitudes and slopes, and the Coulomb
interaction does not enter, and we put $F^C(t)=0$.

The energy dependences of the four quantities are given by 
\begin{equation}  \label{sig-eq}
\sigma(s)= 69.3286+12.6800\log\sqrt{s}+1.2273\log^2 \sqrt{s} ~,
\end{equation}
\begin{equation}  \label{BI-eq}
B_I(s) = 16.2472 + 1.5392 \log\sqrt{s}+0.17476 \log^2\sqrt{s} ~ ,
\end{equation}
\begin{equation}  \label{BR-eq}
B_R(s) = 22.835 + 2.862 \log{\sqrt{s} }+0.32972 \log^2\sqrt{s} ~,
\end{equation}
and 
\begin{equation}  \label{rho-eq}
\rho(s)= \frac{3.528018+0.785609 \log\sqrt{s} } {25.11358+4.59321 \log\sqrt{%
s}+0.444594\log^2 \sqrt{s} } ~ ,
\end{equation}
where $\sqrt{s}$ is in TeV, $\sigma$ in milibarns, $B_I$ and $B_R$ are in $%
\nobreak\,\mbox{GeV}^{-2}$; $\rho$ is dimensionless, passes through a
maximum at about 1.8 TeV, and decreases at higher energies, with asymptotic
value zero. The ratio $B_R/B_I$ is always larger than one, as expected from
dispersion relations \cite{ferreira2}, and behaves asymptotically like 
\begin{equation}
\frac{B_R}{B_I} \rightarrow 1.8867 - \frac{0.24065}{\log\sqrt{s}} - \frac{%
42.6214} {\log^2\sqrt{s} } ~.
\end{equation}

The dimensionless ratio 
\begin{equation}  \label{ratioRI}
R_I = \frac{1}{ (\hbar c)^2 } \frac {\sigma}{16\pi B_I}
\end{equation}
is important in the study of the  form of the pp interaction.  
 In our description of the pp system, as given by the energy
dependences in Eqs. (\ref{sig-eq}-\ref{rho-eq}), this ratio has the high
energy behaviour 
\begin{equation}  \label{ratioRIasymp}
R_I =\frac{1}{ (\hbar c)^2 } \frac {\sigma }{16\pi B_I} \rightarrow 0.3588+ 
\frac{0.54673}{\log\sqrt{s}}-\frac{17.9064}{\log^2\sqrt{s}} ~ .
\end{equation}

From the pure exponential behaviour for small $t$, as we consider in
this paper, this ratio is numerically equal to the ratio $\sigma _{\mathrm{pp%
}}^{\mathrm{el,I}}/\sigma $ between integrated elastic (from the imaginary
part) and total pp cross section. Thus this elastic ratio is also nearly
1/3, and the inelastic ratio is $\sigma _{\mathrm{pp}}^{\mathrm{inel}%
}/\sigma ~\approx ~2/3$ at the highest observed energies . Eq. (\ref%
{ratioRIasymp}) indicates that this ratio converges to $0.3588$ for$\sqrt{s}%
\rightarrow \infty ,$which is far from the value 1/2 that is characteristic
of the idea of a black disk. Our results show that there is no such black
disk behaviour \cite{menon}. This point will be discussed later in $b$-space
representations.

We remark that we have used the slope $B_I$ in the ratio (\ref{ratioRI})
defined above. We may similarly define the ratio using the $B_R$ slope, and
then we obtain the high $\sqrt{s}$ behaviour 
\begin{equation}  \label{ratioRR}
R_R=\frac{1}{ (\hbar c)^2 }\frac { \sigma}{16\pi B_R} \rightarrow 0.1896 + 
\frac{0.325061}{\log\sqrt{s}} - \frac{5.23579}{\log^2{\sqrt{s} } } ~.
\end{equation}
With pure exponential form in the real amplitude, this fraction would be
equal to the ratio $( \sigma_{\mathrm{pp}}^{\mathrm{el,R}}/{\rho^2}) /
\sigma $ . Since $\rho$ is small, the contribution of the real part to the
integrated elastic cross section is also small.

The energy dependence of the two ratios $R_I$ and $R_R$ is shown in Fig. \ref%
{trueratios}.

\begin{figure}[b]
\includegraphics[width=8cm]{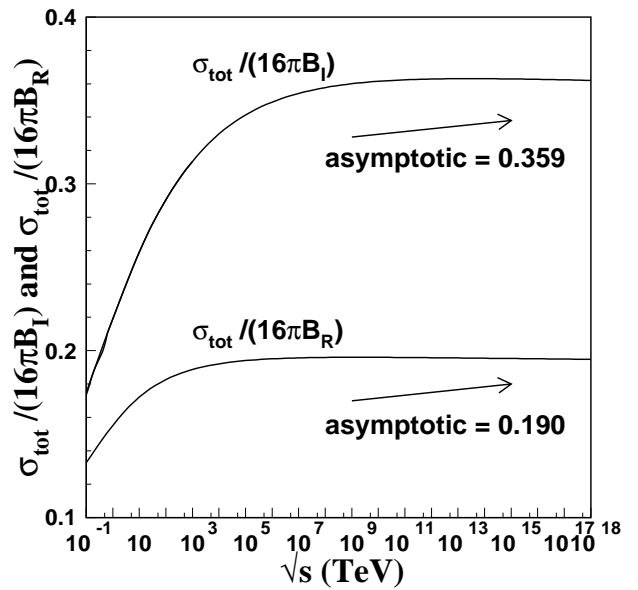}
\caption{ Energy dependence of the dimensionless ratios between total pp
cross section and the slopes $B_I$ and $B_R$, as defined by Eqs. (\protect
\ref{ratioRI}, \protect\ref{ratioRR}). The expressions have finite
asymptotic limits, as shown in equations and in the plots.}
\label{trueratios}
\end{figure}


\section{ p-air Calculations \label{Glauber-section}}

The information on the parameters given above for the pp interaction enters
in the calculation of production cross section $\sigma _{\mathrm{p-air}}^{%
\mathrm{prod}}$ that is obtained from the analysis of EAS in the standard
Glauber approximation \cite{Glauber,Good_Walker, Engel}.  

In the approximation where the forward amplitudes are expressed as pure
exponential form in $t$-space, the corresponding amplitudes in $b$-space
become Gaussians in $b$, 
\begin{eqnarray}
\widehat{T}_{\mathrm{pp}}(s,\vec{b})&=&\widehat{T}_{R}(s,\vec{b})+i\widehat{%
T}_{I}(s,\vec{b})  \nonumber \\
\rightarrow &&\frac{\sigma _{\mathrm{pp}}^{\mathrm{tot}}}{4\pi (\hbar c)^{2}}%
\bigg[\frac{\rho }{B_{R}}e^{-\frac{b^{2}}{2B_{R}}}+i\frac{1}{B_{I}}e^{-\frac{%
b^{2}}{2B_{I}}}\bigg]~.  \label{Tpp}
\end{eqnarray}%
Here, our forward amplitudes $(s,t)$ show different $t$ behaviour in the
imaginary and real parts, with different slopes $B_{I}$ and $B_{R}$. In
terms of the eikonal function $\chi (s,\vec{b})$ this is written 
\begin{equation}
-i~\widehat{T}_{\mathrm{pp}}(s,\vec{b})~=~1-e^{i\chi _{\mathrm{pp}}(s,\vec{b}%
)}~\equiv ~\Gamma _{\mathrm{pp}}(s,\vec{b})~.  \label{iT-eikonal}
\end{equation}%
The term $e^{i\chi _{\mathrm{pp}}(s,\vec{b})}$ represents the S-matrix
function in $b$-space. The optical theorem for pp scattering appears as 
\begin{equation}
\sigma _{\mathrm{pp}}^{\mathrm{tot}}(s)~=~2~(\hbar c)^{2}~\Re ~\int d^{2}%
\vec{b}~\Gamma _{\mathrm{pp}}(s,\vec{b})~.  \label{optical_pp}
\end{equation}

Analogously, for elastic scattering in the p-A system, we define a quantity $%
\Gamma _{\mathrm{pA}}(s,\vec{b})$ that satisfies the optical theorem for the
pA total cross section 
\begin{equation}
\sigma _{\mathrm{pA}}^{\mathrm{tot}}(s)~=~2~(\hbar c)^{2}~\Re ~\int d^{2}%
\vec{b}~\Gamma _{\mathrm{pA}}(s,\vec{b})~.  \label{Sigma_total_H_Ar_2}
\end{equation}
$\Gamma _{\mathrm{pA}}(s,\vec{b})$ is the reaction matrix element averaged
over all target nucleon wavefunction inside the nucleus.

To describe the observed phenomena in EAS, we need to evaluate the quantity 
\[
\sigma _{\mathrm{p-air}}^{\mathrm{prod}}=\sigma _{\mathrm{p-air}}^{\mathrm{%
tot}}-(\sigma _{\mathrm{p-air}}^{\mathrm{el}}+\sigma _{\mathrm{p-air}}^{%
\mathrm{q-el}})~ 
\]%
that is determined experimentally. The quantities named p-air are averages
over a mixture of nitrogen and oxygen nuclei.

Stressing that we provide reliable information on cross sections and
amplitude slopes for the pp scattering input, and a proper, although simple,
treatment of Glauber framework,  our calculations of $\sigma
_{\mathrm{p-air}}^{\mathrm{prod}}$ is adequate for the study of EAS data, 
as shown  in the next section.  

\begin{figure*}[b]
\includegraphics[width=8cm]{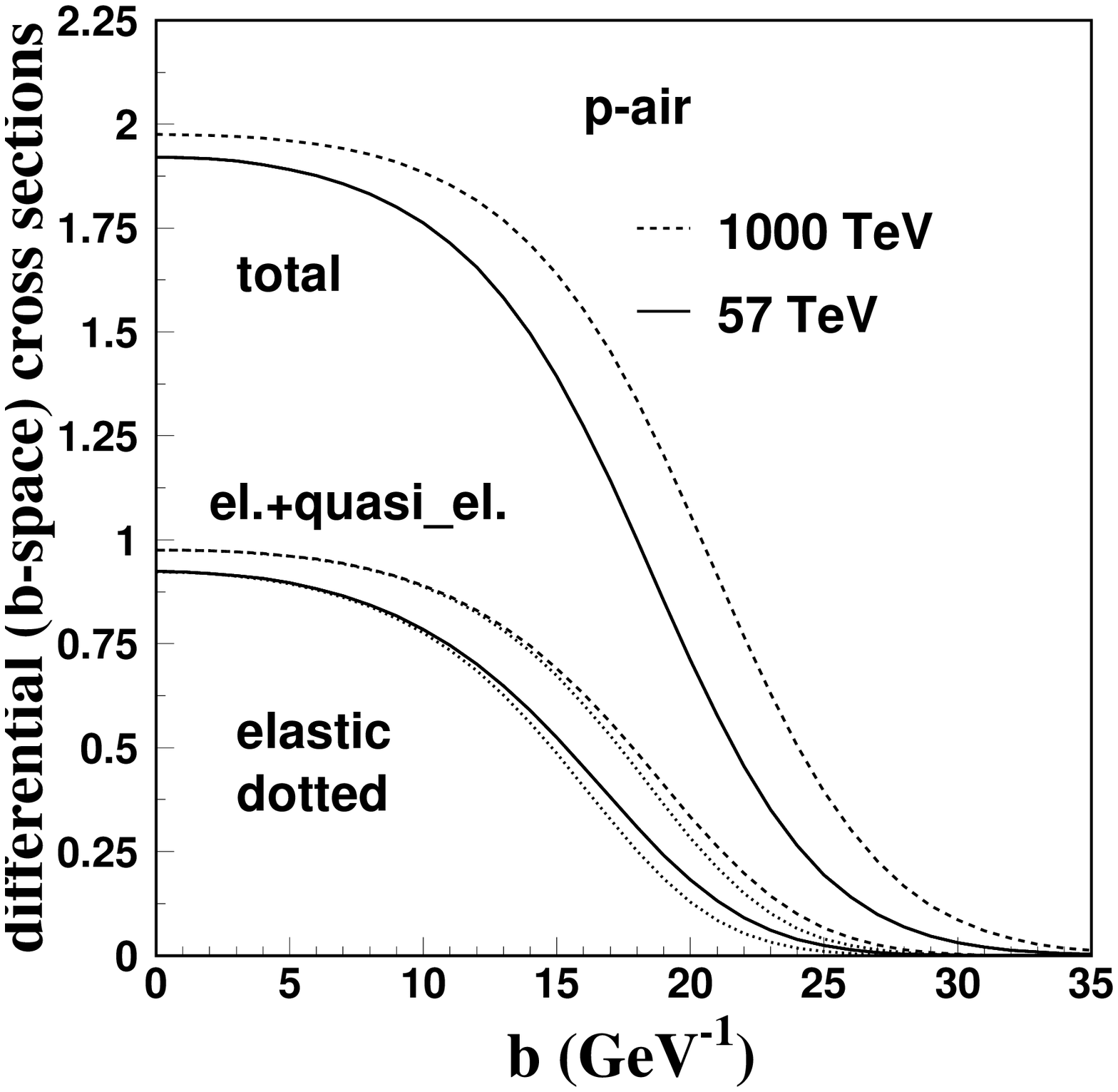} %
\includegraphics[width=8cm]{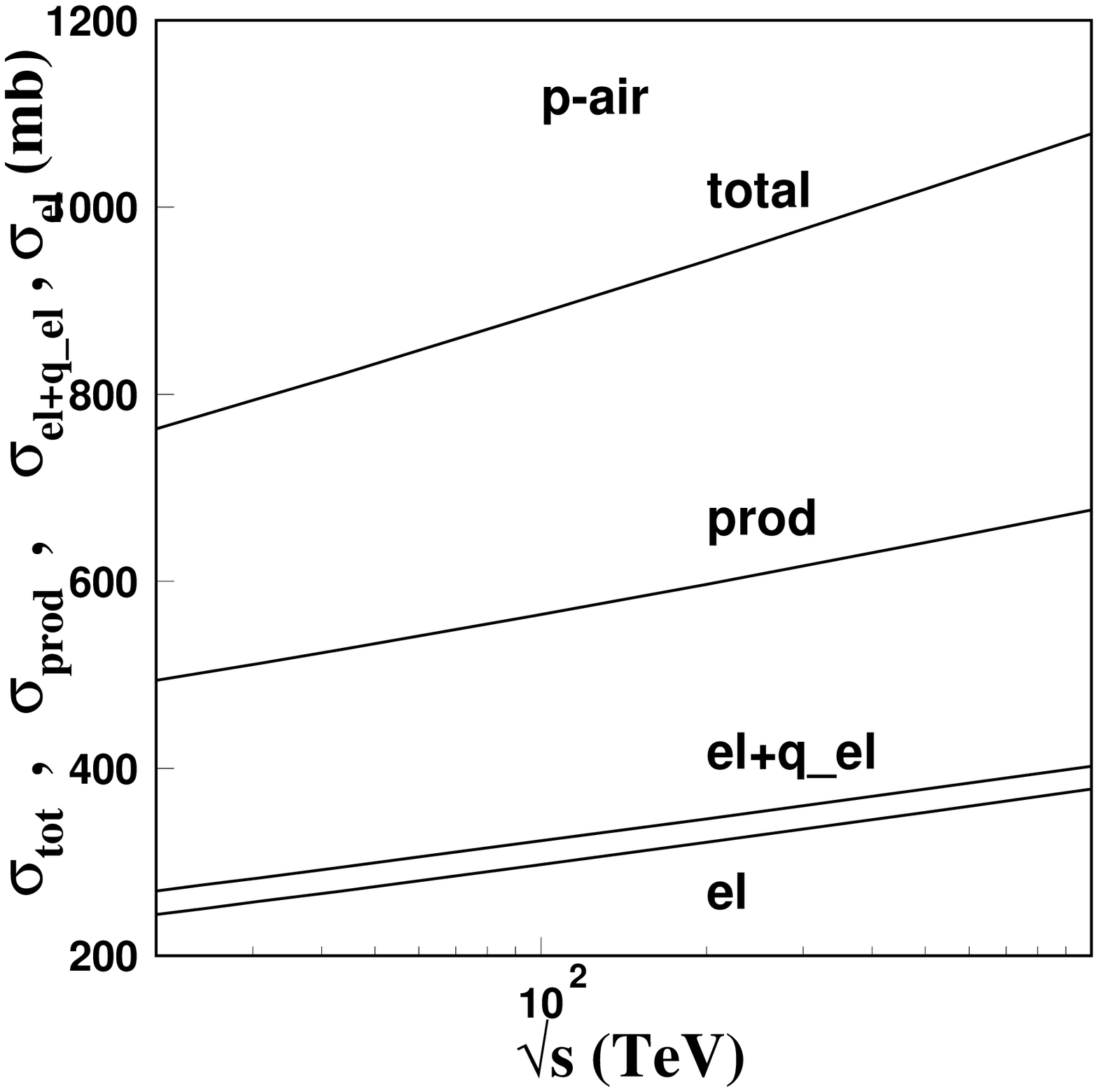}
\caption{ The quantities $d\protect\sigma_{\mathrm{p-air}}^{\mathrm{tot} }/{%
d^{2}\vec{b}} $ and $d\protect\sigma_{\mathrm{p-air}}^{\mathrm{el+q-el} }/{%
d^{2}\vec{b}} $ are plotted as functions of the p-air impact parameter $b$
for the energies 57 and 1000 TeV. As the energy increases, the saturation
limits 2 and 1 are approached by the total and inelastic parts for 
small $b$. The integrated quantities are shown in the second part of the figure. 
The difference between elastic+quasi-elastic and purely elastic terms is small.
}
\label{p_air-fig}
\end{figure*}

The dimensionless quantities that give the $b$-dependence of the total,
elastic+quasi-elastic and pure elastic cross sections for the p-air system
(taking averages over nitrogen and oxygen components) 
\begin{equation}  \label{differential_forms}
\frac{d\tilde\sigma_{\mathrm{p-air}}^{\mathrm{tot}}}{d^{2}\vec{b}} (s,b) ~ , 
\frac{d\tilde\sigma_{\mathrm{p-air}}^{\mathrm{{el+q-el} }}}{d^{2}\vec{b}}%
(s,b) ~ , \frac{d\tilde\sigma_{\mathrm{p-air}}^{\mathrm{el}}}{d^{2}\vec{b}}
(s,b)
\end{equation}
are represented in Fig. \ref{p_air-fig} for the energies $\sqrt{s}=57$ and $%
\sqrt{s}=1000 $ TeV. As in the pp system, the total and inelastic cross
sections for small $b$ approach the limits 2 and 1 as the energy increases.
There is little difference between the elastic+quasi-elastic and the pure
elastic quantities.

The integrated quantities $\sigma_{\mathrm{p-air}}^{\mathrm{tot}} (s),
\sigma_{\mathrm{p-air}}^{\mathrm{el}}+\sigma_{\mathrm{p-air}}^{\mathrm{q-el }%
} (s) ~ $ and $\sigma_{\mathrm{p-air}}^{\mathrm{el}} (s) $ are shown in the
second part of the same figure. The ratio $\sigma_{\mathrm{p-air}}^{\mathrm{%
el}}/\sigma_{\mathrm{p-air}}^{\mathrm{tot}}$ is 0.33 at 57 TeV and 0.35 at
1000 TeV. The difference between elastic+quasi-elastic and purely elastic
contributions is remarkably small, of about 18 \% at 50 GeV and falling
steadily to zero as the energy increases. The inelastic p-air cross section
is about 2/3 of the total, as in the pp system.

\section{ Data from EAS Measurements \label{data-section}}

\begin{figure}[b]
\includegraphics[width=8cm]{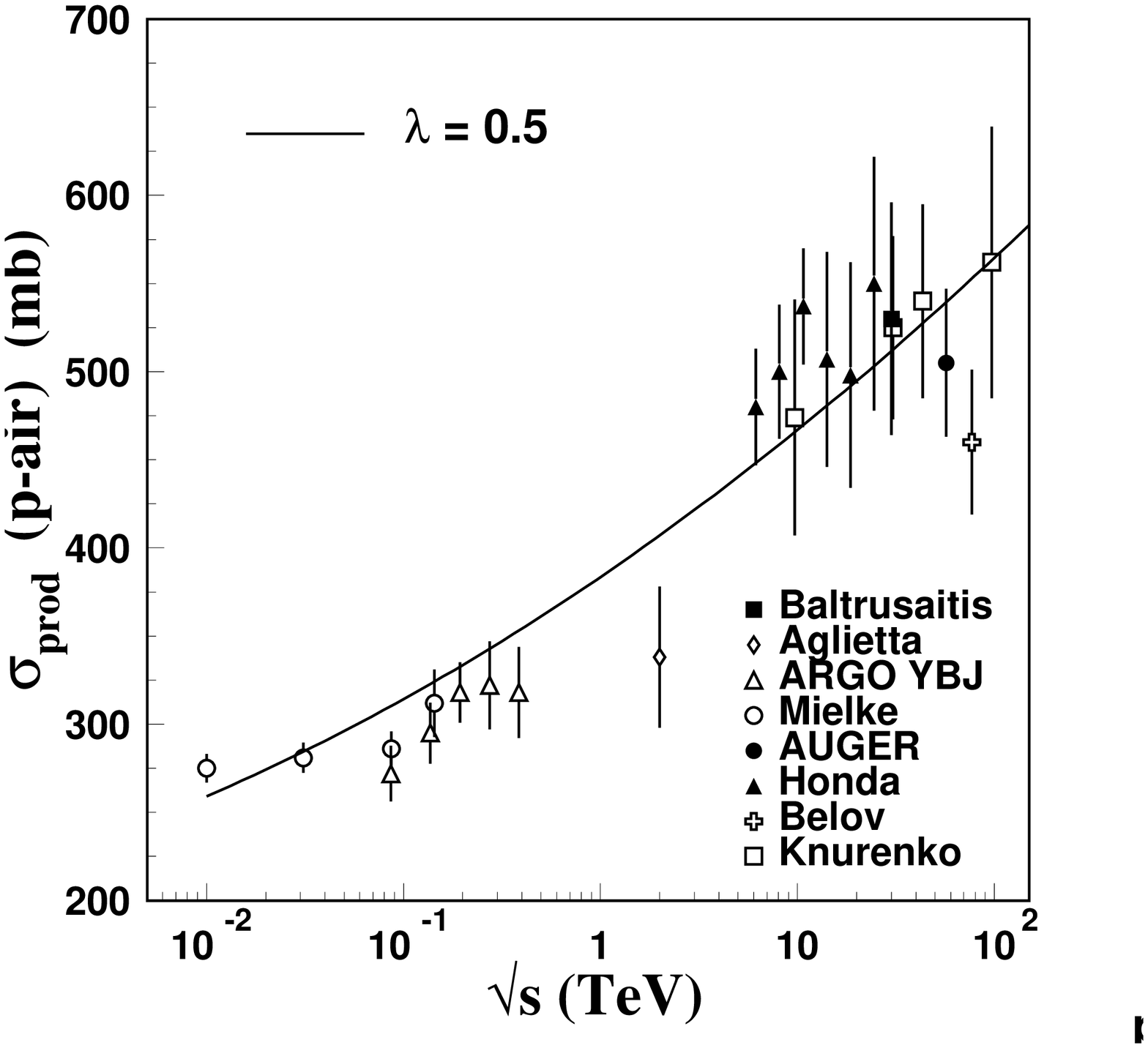}
\caption{ Our calculation of the p-air production cross section is
represented by the solid line, that is well represented by Eq. (\protect\ref%
{curve_data}). The data are from several experiments \protect\cite%
{Auger,Belov,Baltru,Honda, Knurenko,Aielli,Mielke,Aglietta}. Both data and
calculations increase with the energy with a $\log^2 \protect\sqrt{s}$ form. 
}
\label{CR_data-fig}
\end{figure}

Fig. \ref{CR_data-fig} shows our calculation of $\sigma _{\mathrm{p-air}}^{%
\mathrm{prod}}$ with a solid line, together with the data points from EAS
experiments \cite{Auger,Belov,Baltru,Honda, Knurenko,Aielli,Mielke,Aglietta}.

The procedure is straightforward and unique, without free parameters, made
with inputs given by our model for the pp interaction that describes the
elastic differential cross sections at all energies from 20 GeV to 8 TeV in
the whole $t$-range. For the application in Glauber
calculation of the p-air processes, the model enters only in its forward
scattering limit, and is represented by Eqs. (\ref{sig-eq}-\ref{rho-eq}).
Although a pure Gaussian in $b$-space is not adequate for the description 
of the pp amplitudes for large $|t|$, in the Glauber calculation   
the log-squared increases of $\sigma $, $B_{I}$, $B_{R}$ are consequence of
the Yukawa-like behaviour of the exact amplitudes, which do not violate 
unitarity or dispersion relations \cite{KEK-to-be}. Thus we consider that 
this is a reliable input.

The figure shows that in general there is good
agreement between data and 
the calculation of $\sigma _{\mathrm{p-air}}^{\mathrm{prod}}$,  
without any systematic deviation that could require additional terms
  beyond the basic Glauber form.
At high energies above 10 TeV ($\sqrt{s}$ in the proton-proton system) the
agreement is particularly satisfactory within uncertainties of the present
experimental information. In the low energy region we observe that data from
the ARGO-YBJ experiment \cite{Aielli} is below the theoretical curve, while
the data from the Kaskade experiment \cite{Mielke} do not shown the same
systematic deviation.

The theoretical curve for the production cross section can be put in the
simple and convenient form 
\begin{equation}  \label{curve_data}
\sigma_{\mathrm{p-air}}^{\mathrm{prod}}=383.474+33.158 \log \sqrt{s}+1.3363
\log^2\sqrt{s} ~ ,
\end{equation}
with $\sqrt{s}$ in TeV. 

We observe that the data and our calculations of $\sigma_{\mathrm{p-air}}^{%
\mathrm{prod}}$ increase with similar $\log^2 \sqrt{s} $   dependence
as the pp cross sections, but more slowly. To compare the two rates and give
more evidence of regularity in the data, we show in Fig. \ref%
{ratio_sigmas-fig} the relation $\sigma_{\mathrm{p-air}}^{\mathrm{prod}%
}/\sigma(\mathrm{pp) }$ for a set of selected data (chosen by regularity
reasons) together with our calculations. The ratio decreases regularly,
approaching a finite and distant asymptotic limit, as pointed out by the
relation of forms in Eqs. (\ref{curve_data}) and (\ref{sig-eq}). The
existence of a finite asymptotic limit for this ratio and its numerical
value at ultra-high energies is important.

\begin{figure}[b]
\includegraphics[width=8cm]{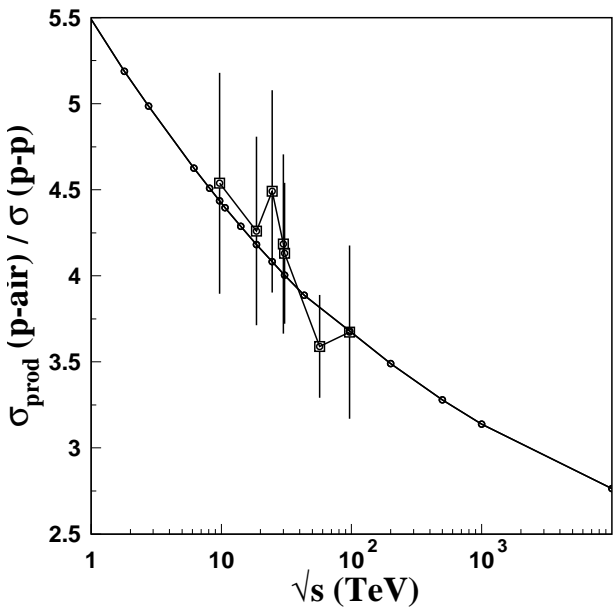}
\caption{ Ratio of p-air and pp cross sections. We show our calculation in
solid line together with selected data \protect\cite%
{Auger,Baltru,Honda,Knurenko}. We observe regular behaviour in the energy
variation of the data for the ratio, that slowly approaches a finite
asymptotic limit. }
\label{ratio_sigmas-fig}
\end{figure}

\begin{figure}[b]
\includegraphics[width=8cm]{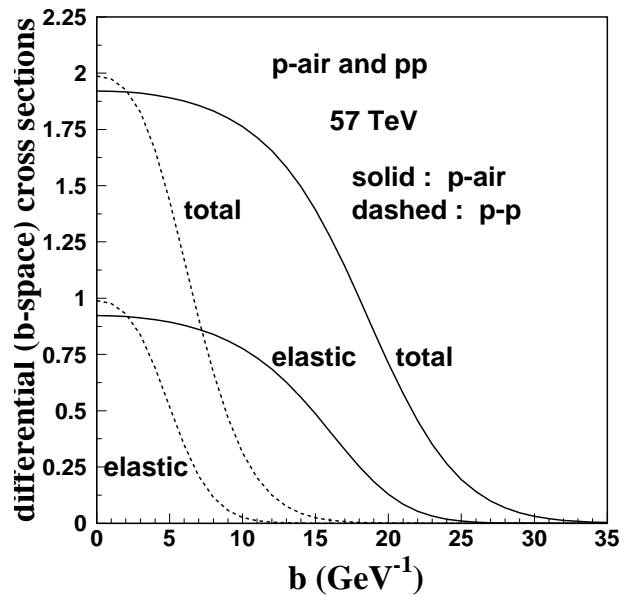}
\caption{ Differential cross sections of the form ${d\tilde{\protect\sigma}}/%
{d^{2}\vec{b}}(s,b)$ (as in Eq. (\protect\ref{differential_forms})) for the
pp and p-air systems at $\protect\sqrt{s}=57$ TeV. The smoothly decaying
tails show that neither the pp nor the p-air systems indicate a black-disk
behaviour. This has consequences for the asymptotic behaviour. 
The  calculation for pp is made with the full amplitudes of our model 
 \cite{KEK_2013,KEK_2013b, KEK-JPhysG2014, KEK-to-be,
KEK_2014}, according to Eq. (\ref{b-amp}).  } 
\label{b_distributions-fig}
\end{figure}

To exhibit more clearly the connection of this experimental behaviour with
the geometric character of the $b$-distributions that build the cross
sections values, we present in Fig. \ref{b_distributions-fig} the forms of
the differential (in $b$-space) cross sections of the pp and p-air systems,
at $\sqrt{s}=57$ TeV. 
At this energy, we confirm that, although the Gaussian approximation fails 
for pp cross sections, for p-air Glauber calculation the effect 
of using pure Gaussian  instead of the full exact amplitudes 
is negligible. 
The distributions show smoothly decaying tails, and their
integrations lead to ratios as $\sigma ^{\mathrm{el}}/\sigma ^{\mathrm{tot}%
}~(\mathrm{pp)}$=0.28, $\sigma ^{\mathrm{el}}/\sigma ^{\mathrm{tot}}~(%
\mathrm{p-air)}$=0.33, $\sigma _{\mathrm{p-air}}^{\mathrm{prod}}/\sigma _{%
\mathrm{pp}}^{\mathrm{tot}}$=3.83, that show that neither the pp nor the
p-air systems tend to a black-disk behaviour, and also that the integrated
distributions are in agreement with the data in Fig. \ref{ratio_sigmas-fig}. 

It seems that, up to this energy domain, the confrontation of our calculation
with data at high energies does not indicate the need of contributions
beyond the standard Glauber calculation. However, the EAS data are not
regular with large error bars, due to uncertainties in the extraction of
values for $\sigma _{\mathrm{p-air}}^{\mathrm{prod}}$. Improvement in the
quality of future data may indicate influence of processes occurring in
intermediate states of the p-air collision, as nucleon excitations,
correlations, shadowing. A particular example is given by the recent AUGER
measurement at 57 TeV, that seems to indicate a lower value with respect to
the general trend of the data, although the error bar is quite large.

In the low energy region, in the data of the ARGO YBJ collaboration \cite%
{Aielli} there may be a regular deviation of our calculations. It may be
that  effects that are not observable at 100 TeV may become important in
this range. Anyhow, the discrepancies are not large, amounting to a maximum
of 10\% : at $\sqrt{s}=0.0865$ TeV the ARGO YBJ experiment gives $\sigma _{%
\mathrm{p-air}}^{\mathrm{prod}}=272\pm 15.8$ mb , while the theory gives
307.21 mb. On the contrary, at $\sqrt{s}=0.031$ TeV the Kaskade experiment 
\cite{Mielke} and the theoretical value coincide very well (at $281\pm 8.5$
and $286$ mb respectively).

In general, there seems to be more room for improvement in the measurements
than in the theoretical calculation, and we believe that our pp input
together with the basic Glauber calculation have successfully passed the
test in the comparison with EAS data.

\begin{figure*}[b]
\includegraphics[width=8cm]{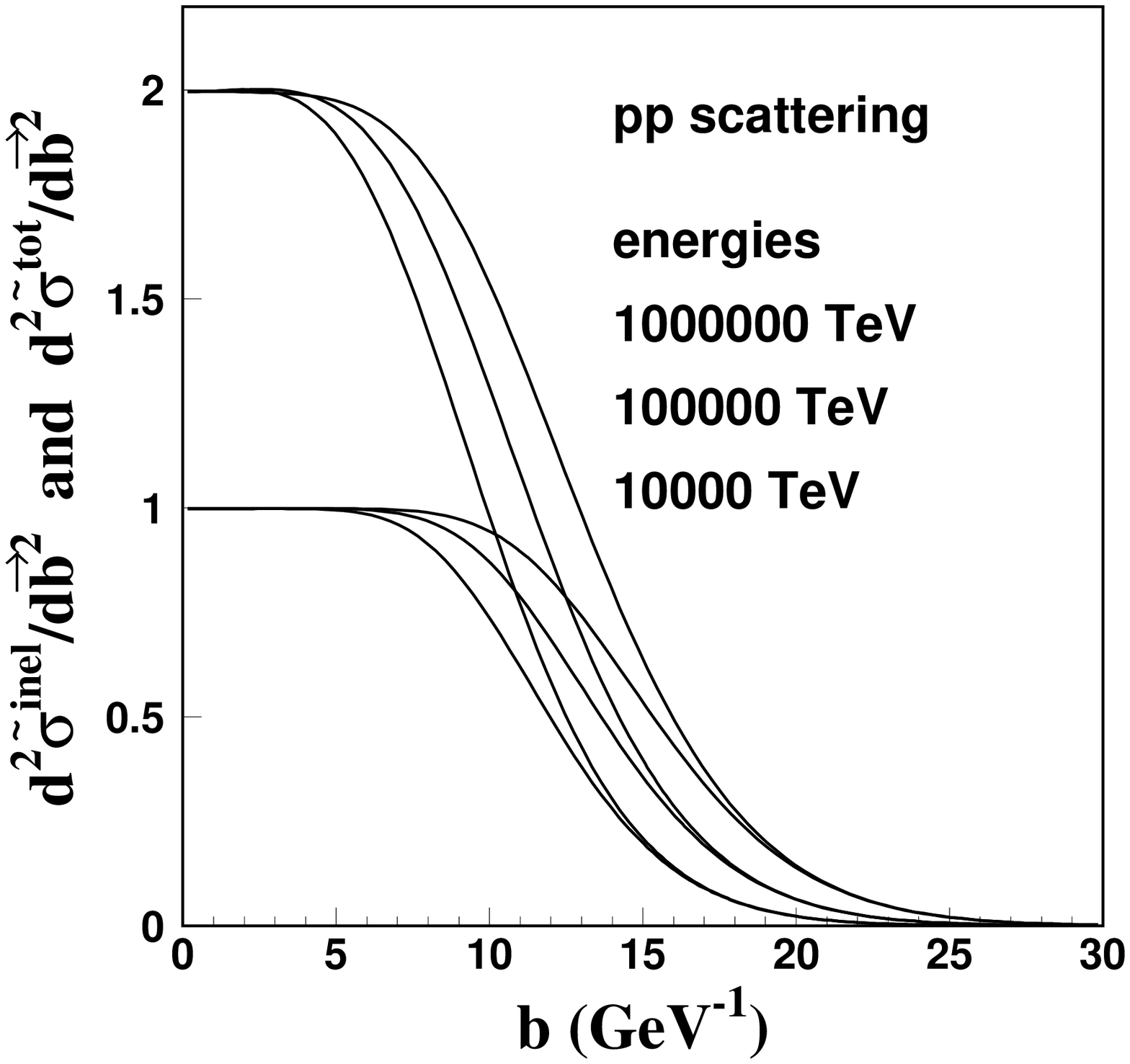} %
\includegraphics[width=8cm]{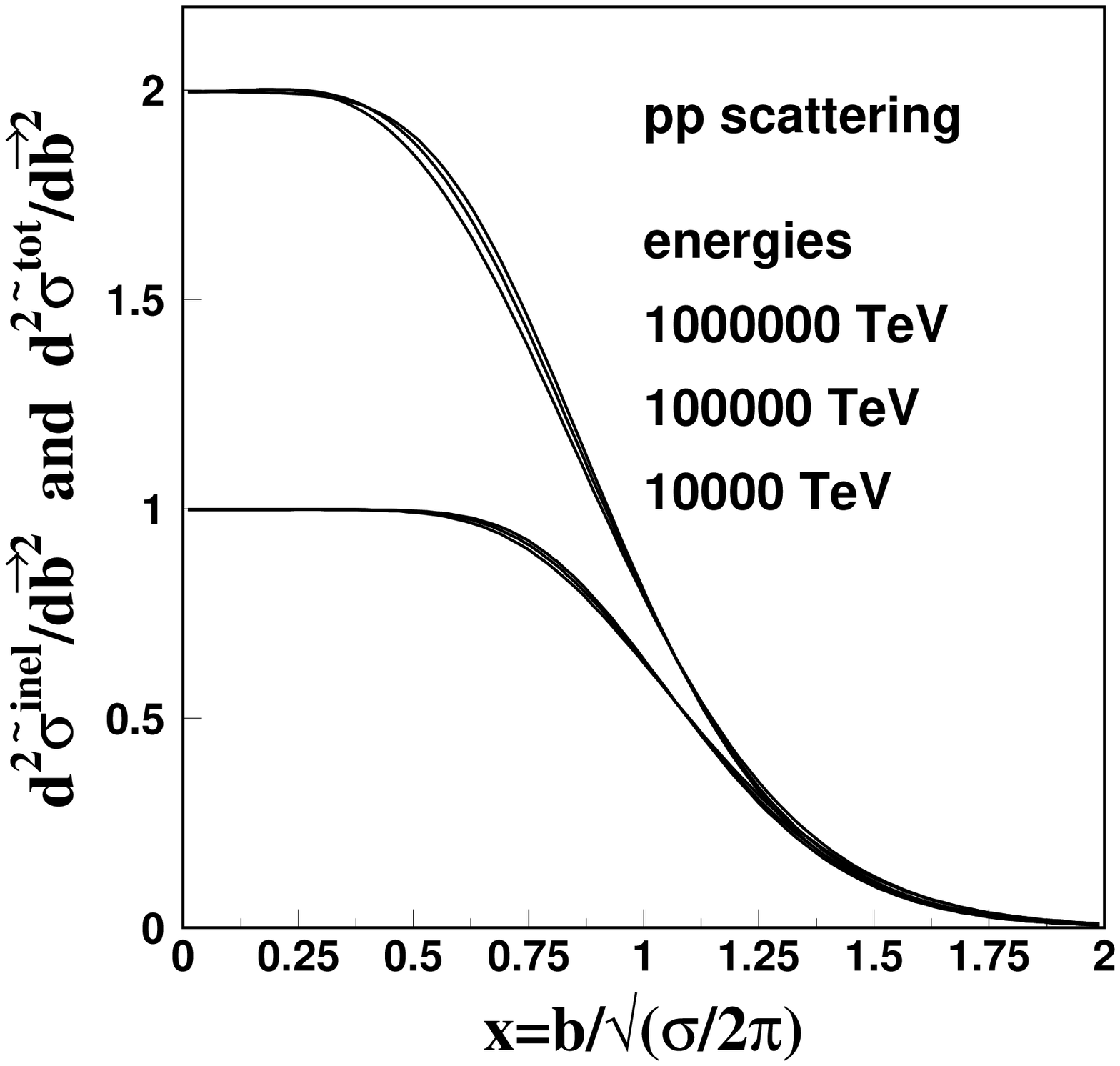}
\caption{Dimensionless differential $b$-space cross sections for total and
inelastic pp interactions. The plotted energies are $10^4$, $10^5$ and $10^6$
TeV. In the second part of the figure, the cross sections are plotted
against the scaled variable $x$, showing universal behaviour. }
\label{scaling-fig}
\end{figure*}

\begin{figure}[b]
\includegraphics[width=8cm]{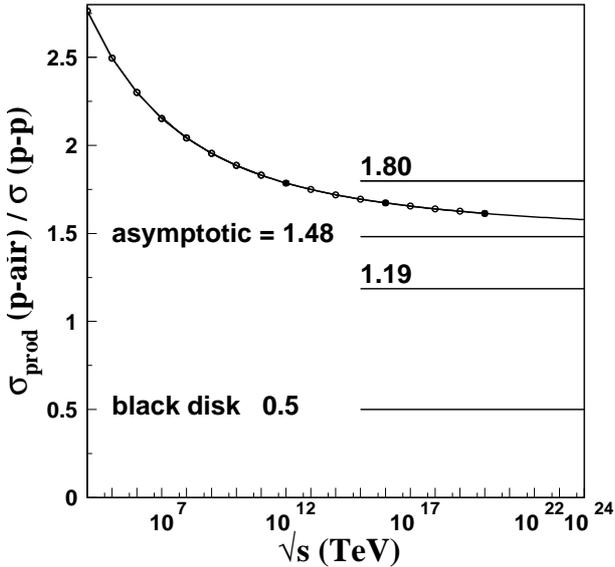}
\caption{ Ratio of p-air and pp cross sections at ultra-high energies.
Calculations are marked with dots and connected with a continuous line. The
dashed line is given analytically by the fraction of $\log ^{2}$ forms for $%
\protect\sigma _{\mathrm{p-air}}^{\mathrm{inel}}(s)$ and $\protect\sigma _{%
\mathrm{pp}}^{\mathrm{tot}}(s)$, given in the text. It gives good
representation of the points for energies above $10^{6}$ TeV and tends to
the asymptotic limit 1.48 ~. As we see from this figure, the asymptotic
value is attained only for really large $\protect\sqrt{s},$ say $%
\protect\sqrt{s}\gg 10^{20}$ TeV. }
\label{asymptotic_ratio-fig}
\end{figure}

\section{ Properties of b-Space Representation}

Although the $b$-space representation  does not exactly mean the
classical impact parameter space, and is not observable, it gives us a
geometrical image of the behavior of a proton from the amplitude. In the
forward amplitude, the corresponding $b$-space is a Gaussian form of the
eikonal thickness function. The complete form of Eq. (\ref{b-amp}) for 
the pp scattering amplitude in $b$-space is 
\begin{eqnarray}
&&\hat{T}_{K}(s,\vec{b}) =\frac{\alpha _{K}}{2\beta _{K}}e^{-b^{2}/4\beta
_{K}}  \nonumber \\
&&+\lambda _{K}\frac{2e^{\gamma _{K}-\sqrt{\gamma _{K}^{2}+b^{2}/a_{0}}}}{%
a_{0}\sqrt{\gamma _{K}^{2}+b^{2}/a_{0}}}\Big[1-e^{\gamma _{K}-\sqrt{\gamma
_{K}^{2}+b^{2}/a_{0}}}\Big]~,  \label{b-amp}
\end{eqnarray}%
where $K=R,I$ indicates either the real or the imaginary part and $%
a_{0}=1.39~\mathrm{{GeV}^{-2}}$ is a constant from the stochastic vacuum
model \cite{dosch,ferreira1}. The second term comes from the stochastic
vacuum model, representing the effect of interaction of Wilson loops. Note 
that due to the presence of this term, the first Gaussian term does not coincide
with that used in the previous section. That is, the slope parameters $%
B_{R,I}$ are not given by $\beta _{K}$'s above but defined as $B_{K}\equiv
d\ln T_{K}^{2}/dt$ which contains contributions from the two terms.
Furthermore, the large $b$ behavior of the amplitudes are not Gaussian
anymore, but falls down with a Yukawa-like tail, $\sim e^{-b/b_{0}}/b$.  We
observe also a very nice geometrical scaling property in differential cross
sections in $b$-space. At asymptotic high energies, they scale with the
variable $x=b/\sqrt{\sigma _{T}\left( \sqrt{s}\right) }$,  as shown in
Fig. \ref{scaling-fig}. As seen clearly from this figure, the profile
functions (differential cross sections in $b$-space) as function of $x$
degenerate for all high energies to respective unique distribution, and
their forms are much different from that expected for the black disk (sharp
edged Heaviside step function). In fact we can show that this diffused
nature of the proton profile functions is responsible for the deviations of
the ratios $R_{I}$\, $\sigma _{\mathrm{pp}}^{\mathrm{el,I}}/\sigma $ and $%
\sigma _{\mathrm{pp}}^{\mathrm{inel}}/\sigma $ from the black disk value 1/2 
  \cite{KEK-JPhysG2014}.   

   It is important to note that the Gaussian approximation for pp
amplitudes at these energies (including that of 
Fig. \ref{b_distributions-fig}) violates  
unitarity so that these curves are calculated with the full exact amplitude, 
 Eq.(\ref{b-amp}). The non-Gaussian behavior of the shape function 
has been pointed out also in \cite{Tamas,Tamas2}.

\section{Final Analysis and Expectations \label{final}}

The amplitudes that we have constructed to describe accurately the pp
elastic differential cross sections at energies from 20 GeV to 8 TeV are
used in Glauber formalism to evaluate the p-air production cross section in
the energy domain of EAS/CR experiments. Our prediction for the whole energy
interval from 10 GeV to 100 TeV of p-air production cross section is shown
in Fig. \ref{CR_data-fig}.

The calculations with Glauber approach depend crucially on the input values
of $\sigma_{\mathrm{pp}}^{\mathrm{tot}}(s)$ and $B_I(s)$, and thus the
results obtained for the high energies of the CR experiments 
are important tests of the energy dependences that we propose for these
quantities, given in Eqs. (\ref{sig-eq}, \ref{BI-eq}). It is particularly
remarkable that the $\log^2$ dependence that we propose for $B_I(s)$
predicts higher values for the extrapolated values of this quantity, and the
data seem to be consistent with this. Thus at 57 TeV we have $B_I= 25 ~%
\nobreak\,\mbox{GeV}^{-2}$, value that is higher than the usual obtained,
for example from Donnachie-Landshoff or Regge form. The comparison with CR
data helps to test such alternatives.

The extraction of fundamental information on the energy dependence of pp
total cross section from CR/EAS measurements depends on this point. Thus our
prediction for pp cross section at 57 TeV is $\sigma$ pp = 140.7 mb. In the
experimental paper \cite{Auger}, where the measured value for $\sigma_{%
\mathrm{p-air}}^{\mathrm{prod}}$ is below our calculation (see Fig. \ref%
{CR_data-fig}), and other theoretical models for $\sigma(s)$ and $B_I(s)$
are used, the reported value for $\sigma$ is $\sigma (\rm pp)= 133 \pm 29$ mb. 
Hopefully this important question will be investigated in future CR measurements.

From our representation of the scattering amplitudes we can calculate the
asymptotic values of quantities that approach finite values at high
energies. These values are important for the geometric interpretation of the
dynamics, as can be studied in the representation of the impact parameter $b$%
. For example, the behaviour of the ratios $\sigma_{\mathrm{pp}}^{\mathrm{tot%
}}/B_I$ and $\sigma_{\mathrm{pp}}^{\mathrm{tot}}/B_R$ are connected with
integrated elastic pp cross sections and thus with the rate of inelastic
processes at high energies in the pp system. Our results show that $\sigma_{%
\mathrm{pp}}^{\mathrm{inel}}/\sigma_{\mathrm{pp}}^{\mathrm{tot}}\sim \approx
2/3 $ at very high energies. This ratio and $R_I$ together show that the
geometric nature of pp cross section is far the black disk form. This is so
even asymptotically.

To acquire a better feeling about the regularity of the energy dependence of
the data and its representation by the theoretical calculation, we present
in Fig. \ref{ratio_sigmas-fig} results on the ratio between p-air and pp
cross sections. The figure shows that this ratio has the important property
of approaching a finite value for infinite energy. This information if of
fundamental importance for the understanding of the geometric nature of the
pp interaction and its energy dependence. We show that this ratio is
intimately related with the relation 
$\sigma({\rm pp inelastic})/\sigma({\rm pp total})$  and
with the behaviour of the eikonal functions for large b.

The important question of the energy dependence of the ratio of p-air to pp
cross sections is studied in a direct way, using properties of the $b$
dependence of pp interaction at high energies. We show that the Yukawa-like
behaviour of the interaction range, inspired in the stochastic vacuum model,
explains quantitatively   the value of the asymptotic limit
of the ratio $\sigma _{\mathrm{p-air}}^{\mathrm{inel}}/\sigma _{\mathrm{pp}%
}^{\mathrm{tot}}$.

 In this work, we have applied the standard Glauber model, 
considering that the nucleons are the scattering centers. Of course, for a
ultra-high energy domain, where the pp cross section  becomes
comparable to  the geometric cross section of a target nucleus, the Glauber
approach itself may be questionable. In the Glauber approach of pA cross
section, the scattering centers inside the target are nucleons, with a fixed
distribution determined by the nuclear wave function. However, at the
energies where the interaction size of pp becomes large enough so that their
superposition becomes not negligible, the scattering centers are rather
partons and not nucleons. Then the energy dependence of pA cross section can
become drastically different \cite{licinio} . Here we have an open question.
Further theoretical investigations of microscopic structures leading to the
asymptotic behaviour in p-air cross sections will be very interesting.

\begin{acknowledgement}
The authors wish to thank the Brazilian agencies CNPq, CAPES, PRONEX and 
FAPERJ for financial support.  
\end{acknowledgement}

\end{document}